\documentclass[aps,preprint,superscriptaddress]{revtex4-2}
\usepackage{graphicx}
\usepackage{epstopdf}
\usepackage{color}
\usepackage{bm}
\usepackage{amsmath}
\usepackage{changes}
\usepackage[colorlinks=true,linkcolor=blue,urlcolor=blue,citecolor=blue]{hyperref}

\begin{document}

\title{Macroscopic bioinspired magnetic active matter and the physical limits of magnetotaxis}

\author{N\'estor Sep\'ulveda}
\affiliation{School of Engineering and Sciences, Universidad Adolfo Ib\'a\~nez, Diagonal las Torres 2640, Pe\~nalol\'en, Santiago, Chile.}

\author{Francisca Guzm\'an-Lastra}
\affiliation{Department of Physics, Universidad de Chile, Santiago, Chile.}

\author{Miguel Carrasco}
\affiliation{Universidad Diego Portales, Facultad de Ingenier\'ia y Ciencias, Santiago 8370191, Chile.}

\author{Bernardo Gonz\'alez}
\affiliation{School of Engineering and Sciences, Universidad Adolfo Ib\'a\~nez, Diagonal las Torres 2640, Pe\~nalol\'en, Santiago, Chile.}
\affiliation{Center of Applied Ecology and Sustainability (CAPES), Santiago, Chile.}

\author{Mariana Navarro}
\affiliation{Condensed Matter i-Lab, Universidad Adolfo Ib\'a\~nez, Diagonal las Torres 2640, Building D, Pe\~nalol\'en, Santiago, Chile.}

\author{Eugenio Hamm}
\affiliation{Departamento de F\'isica, Universidad de Santiago de Chile, Av. V\'ictor Jara 3493, Estaci\'on Central, Santiago, Chile.}

\author{Andr\'es Concha}
\email{andres.physics.research@gmail.com}
\affiliation{School of Engineering and Sciences, Universidad Adolfo Ib\'a\~nez, Diagonal las Torres 2640, Pe\~nalol\'en, Santiago, Chile.}
\affiliation{Condensed Matter i-Lab, Universidad Adolfo Ib\'a\~nez, Diagonal las Torres 2640, Building D, Pe\~nalol\'en, Santiago, Chile.}
\affiliation{CIIBEC, Research Center, Santiago, Chile.}

\begin{abstract}
Magnetotactic bacteria (MTB) are endowed with an exquisite orientation mechanism allowing them to swim along the geomagnetic field lines. This mechanism consists of a chain of bio-synthesized magnetic nano-crystals that endow MTB with a permanent magnetic moment. Although the physics behind the minimum size of this biological compass is well understood, it is yet unclear what sets its maximum size. Here we combine macroscopic bioinspired experiments, calibrated simulations, and analytic estimates to show that increasing dipolar strength can drive magnetic active matter from a freely swimming regime into clustered states. Using a physical model with parameters relevant to MTB, we infer a plausible physical upper bound on useful magnetosome-chain magnetic moments: beyond a threshold, clustering and the formation of compound bodies are expected to hinder effective swimming and reduce magnetotactic performance. Our macroscopic bio-inspired experiment and physical model show how long-range  magnetic interactions reshape the phase behavior of anisotropic active matter and provide a programmable platform for studying magnetic active matter across scales.
\end{abstract} 
\maketitle
\begin{figure}[t!] 
\includegraphics[width=0.90\textwidth,angle=0,clip=true,trim=0 0 0 0]{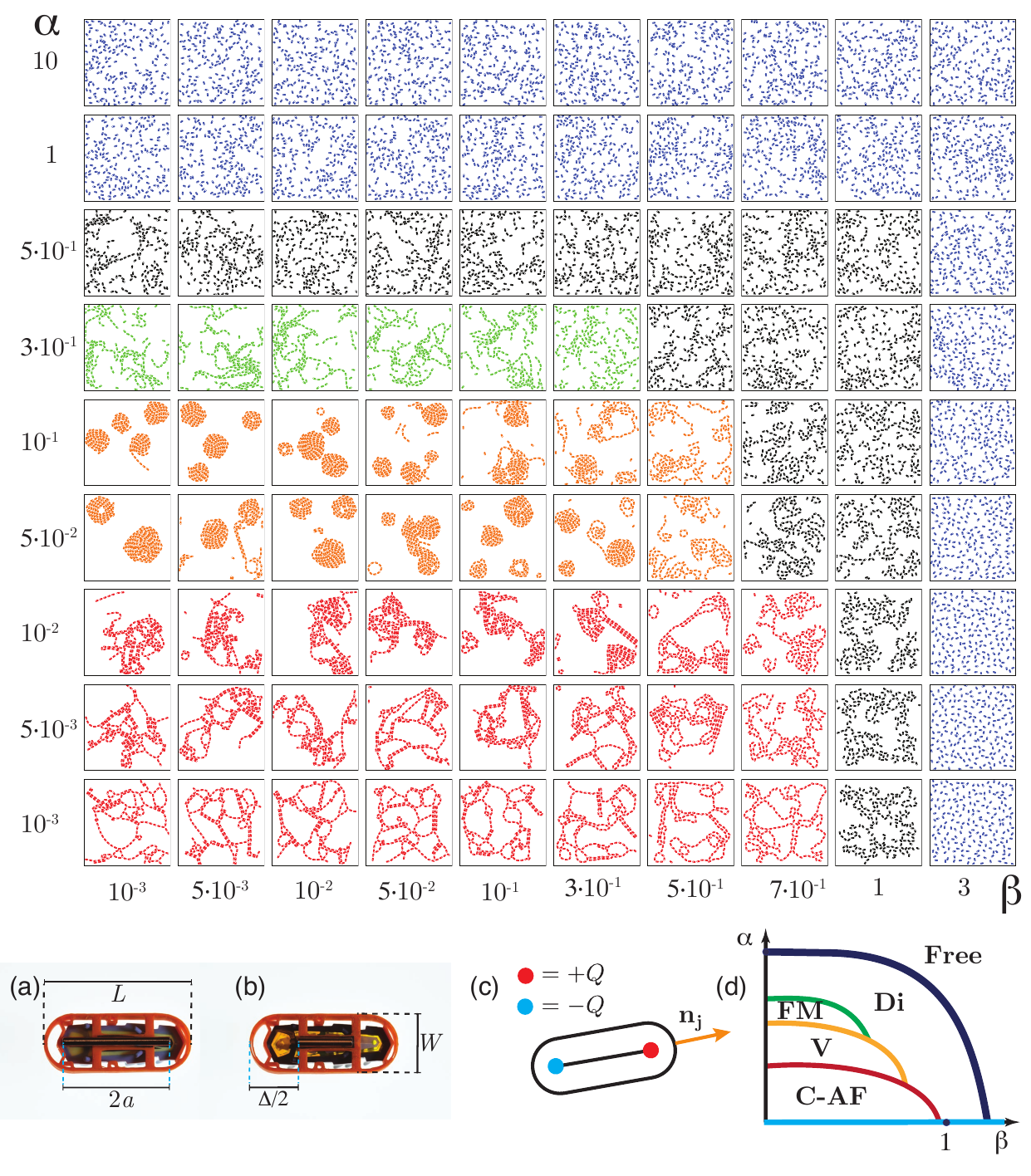}
\caption{Magnetic Active Matter (MAM) and its phase diagram. (\textbf{a,b}) Hexbug Nano robots fitted with a 3D-printed armor and Neodymium magnets of two different lengths, $2a=38.1$ mm and $19.05$ mm; the total body length is $L=51.8$ mm in both cases. The armor holds the magnets and provides a smooth hard boundary for collisions. Each cylindrical magnet has radius $r=1.59\times10^{-3}$ m and saturation magnetization $M_s=1.1\times10^{6}\,\mathrm{A/m}$ (see Supplementary Material S5). (\textbf{c}) Dumbbell representation of a magnetic rod, with pole charges $+Q$(Red) and $-Q$(Blue) at its ends, where $Q=M_s\pi r^2$ \cite{mellado2012macroscopic,concha2018PRL}. $\Delta$ is the shortest distance between two collinear magnets, and $\hat n_i$ is the director of particle $i$. See Supplementary Movies 1-6. (\textbf{d}) Schematic phase diagram in terms of the dimensionless parameters $\alpha$ and $\beta$, which compare inertia with activity and rotational noise with magnetic interactions, respectively. The observed regimes are Free (noninteracting), Di (dimers), FM (winding/unwinding magnetic vortices with local ferromagnetic order), V (stable vortices), and C-AF (antiferromagnetic clusters). The line $\alpha=0$ (cyan) corresponds to the regime relevant for natural or artificial microswimmers.} 
 \label{FIG1}
\end{figure}

Magnetotactic bacteria (MTB) contain magnetosomes, which are intracellular, ferrimagnetic nano-crystals of magnetite or
greigite. Magnetosomes are usually arranged in a linear chain within each bacterium (e.g: {\it{Magnetospirillum magneticum}}, {\it{Magnetospirillum magnetotacticum}}) \cite{blakemore1975magnetotactic,frankel1980navigational,dunin1998magnetic,schuler1999bacterial,komeili2006magnetosomes}, with few known exceptions ({\it{Magnetobacterium bavaricum}}) \cite{jogler2010cultivation}. This linear chain provides a net magnetic moment, $\vec{M}$, that allows MTB to align with the geomagnetic field, $\vec{H}$, in a noisy environment \cite{frankel1980navigational}. This is because oxygen is toxic for these microaerophilic or anaerobic organisms, and it is beneficial for them to be directed toward and kept in the sediments away from the water surface  where oxygen is abundant \cite{blakemore1975magnetotactic,smith2006quantifying}. This mechanism can be understood as the competition between the aligning  dipolar energy, $\vec{M}\cdot \vec{H}$, and the thermal noise $k_{B}T$. Thus, for MTB to align with the Earth?s magnetic field, both quantities should be comparable. This well known and accepted criterion provides a lower bound for the magnetosome chain size \cite{frankel1980navigational,esquivel1986motion,klumpp2016magnetotactic}. 
In the same spirit, we should expect that the larger the dipolar moment of MTB, the better. However, MTB with large magnetic moments have not been observed in nature. Inspired by this observation, and the general subject of the effect of long range interactions in active matter, we have built an artificial system that allows for the exploration of these questions, and provide a flexible tabletop experiment to investigate magnetic active matter (MAM), Fig.\ref{FIG1}. 

\begin{figure}[ht!] 
\includegraphics[width=0.95\textwidth,angle=0,clip=true,trim=0 0 0 0]{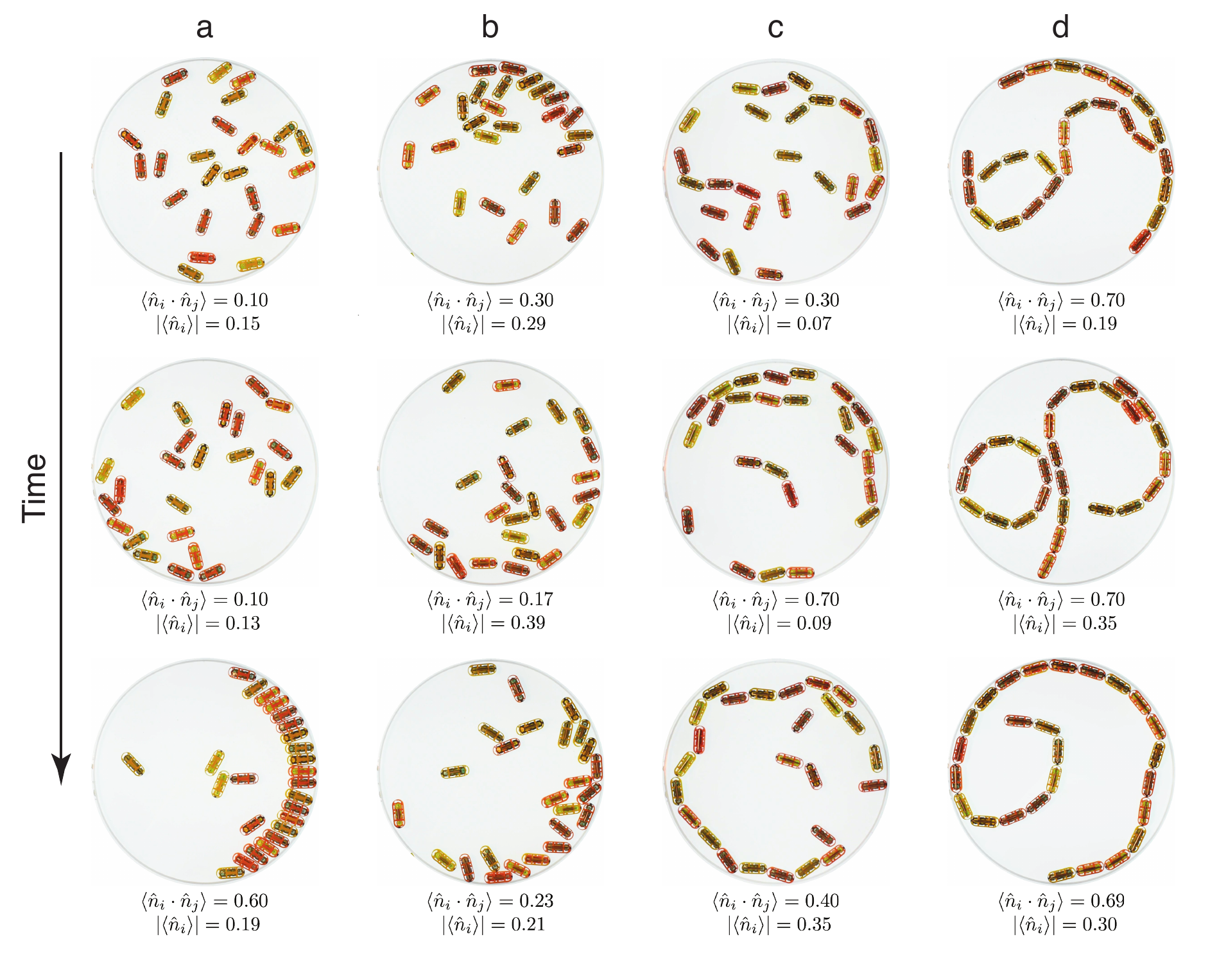}
\caption{Experimental configurations and correlations of magnetic active matter in a domain of diameter $400$ mm. (\textbf{a}) Zero magnetic interactions. (\textbf{b}) MagD-bots with magnets of length $2a=19.05$ mm and diameter $2r=3.175$ mm, with $Q=7.9$ Am and $N=25$. (\textbf{c}) Same as (\textbf{b}), but with $2a=38.10$ mm. (\textbf{d}) Same as (\textbf{b}), but with $2a=44.45$ mm and $N=26$. Below each panel we show the magnitude of the average director and the correlation $\langle \hat n_i \cdot \hat n_j\rangle$. See also Supplementary Movies 1-6.
} 
\label{FIG2}
\end{figure}

\begin{figure}[!t] 
\includegraphics[width=0.95\textwidth,angle=0,clip=true,trim=0 0 0 0]{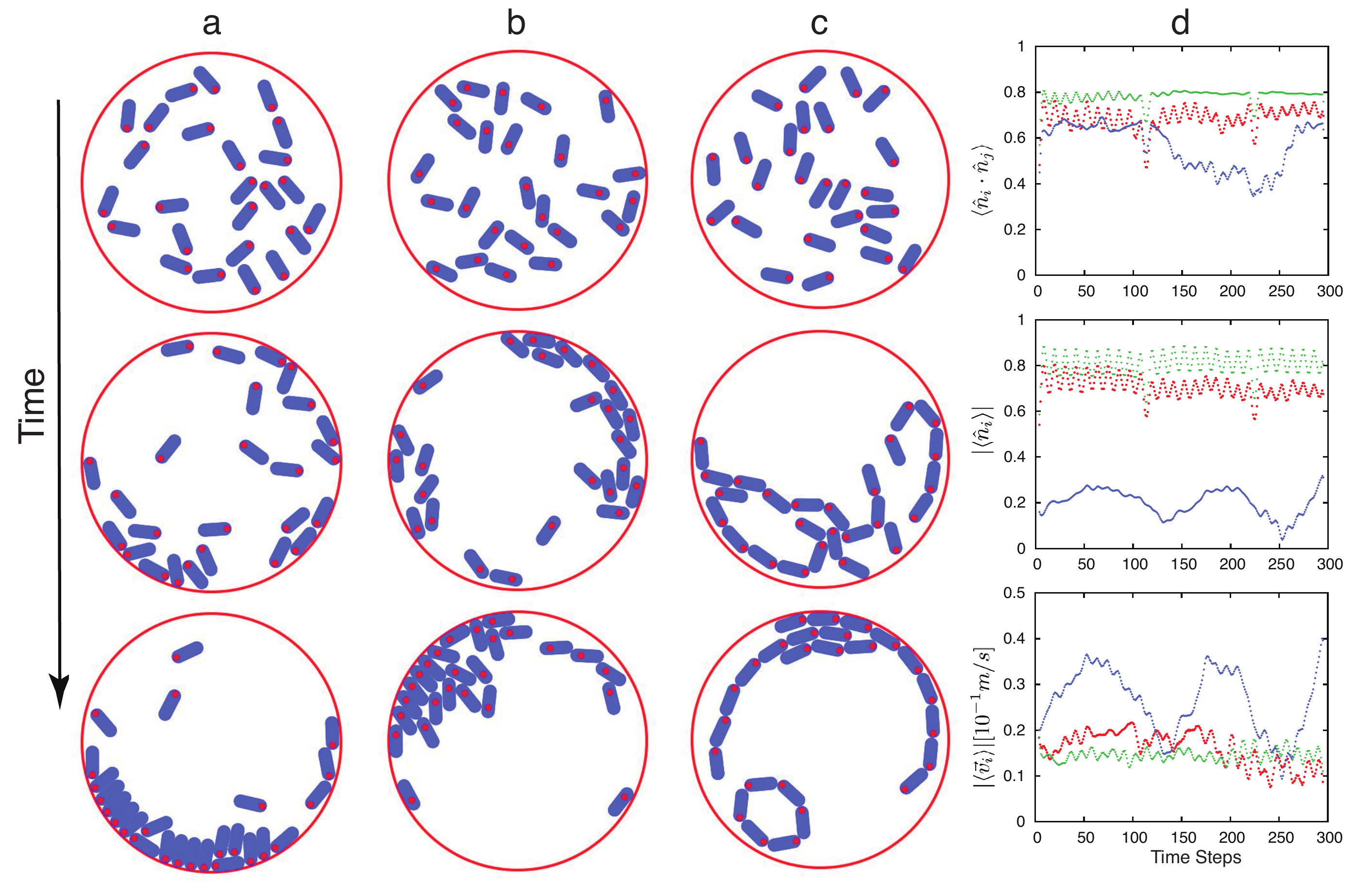}
\caption{Numerical configurations and correlations of magnetic active matter for $N=25$ MagD-bots in a $400$ mm domain. Simulations start from a random initial condition at $t=0$. (\textbf{a}) $Q=0$. (\textbf{b}) $Q=8$, $2a=19.05$ mm. (\textbf{c}) $Q=8$, $2a=38.10$ mm. All other parameters are fixed. (\textbf{d}) From top to bottom: $\langle \hat n_i\!\cdot\!\hat n_j\rangle$, $|\langle \hat n_i\rangle|$, and $|\langle \vec v_i\rangle|$ versus time for (\textbf{a})-(\textbf{c}); red dots, green triangles, and blue rhombi denote (\textbf{a}), (\textbf{b}), and (\textbf{c}), respectively. See Supplementary Movies 7-9.} 
\label{FIG3}
\end{figure}

\begin{figure*}[!t] 
\includegraphics[width=0.90\textwidth,angle=0,clip=true,trim=0 0 0 0]{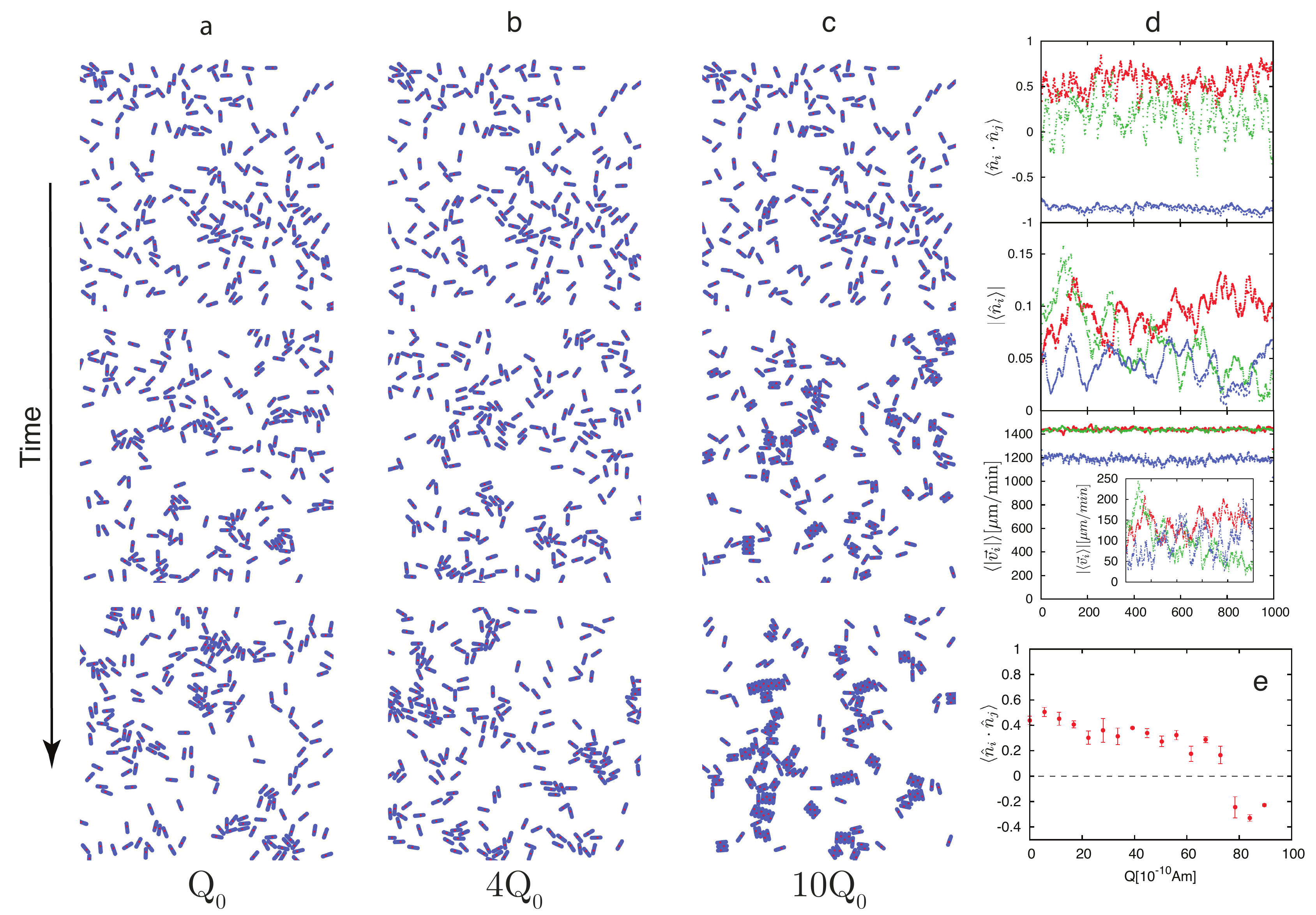}
\caption{Numerical simulations in the parameter regime relevant for MTB. (\textbf{a}) For $Q_0=9.4\times10^{-10}\,\mathrm{Am}$, $2a=1.1\,\mu\mathrm{m}$, $L=5\,\mu\mathrm{m}$, $W=1.6\,\mu\mathrm{m}$, and $M\sim10^{-15}\,\mathrm{kg}$, MTB do not self-assemble into superstructures. (\textbf{b}) Same parameters as in (\textbf{a}), but with increased magnetic charge $Q=4Q_0$. (\textbf{c}) For $Q=10Q_0$, the system rapidly forms antiferromagnetically ordered clusters that hinder free swimming. (\textbf{d}) From top to bottom: nearest-neighbor correlations, average spin, and average velocity (inset: average speed) versus time for cases (\textbf{a})-(\textbf{c}), shown as red dots, green triangles, and blue rhombi, respectively. See also Supplementary Movies 10--12. (\textbf{e}) Nearest-neighbor correlations versus magnetic charge $Q$, showing a transition from ferromagnetic to antiferromagnetic order at $Q=7.5\times10^{-9}\,\mathrm{Am}$. 
}
\label{FIG4}
\end{figure*}

In this article, we argue that there is a strong physical constraint for the maximum dipolar moment of single magnetosome chains. This constraint arises from the fact that MTB need to be able to swim into regions of low oxygen, but for large dipolar moments a collection of MTB will enter into a phase where they will form magnetic clusters making it impossible for them to efficiently swim (See Figs.\ref{FIG1}-\ref{FIG4}). Thus, if  mutants grow magnetosome chains with extremely large magnetic moments they will self-assemble into dimers, or magnetic clusters (See Figs. \ref{FIG1}, \ref{FIG3} and \ref{FIG4}), generating a new compound object in which flagellar motion will be suppressed, decreasing their ability to swim. Thus, MTB with extremely large magnetic moments will not prevail.

The generic formation of clusters of particles being them inanimate or active shows up in a plethora of phenomena ranging from chemistry to physics, biology to astrophysical objects \cite{whitesides2002self,whitesides2002beyond,ong2017programmable,makey2020universality,lee2015direct,kravtsov2012formation,balbus1998instability}. The existence and evolution of these structures at different length scales are intimately tied to the type of interactions between their basic constituents, particle symmetries, and emergent symmetries that may show up due to interactions \cite{anderson1972more,huber2018emergence,denk2020pattern}.  For active agents endowed only with self-avoidance and/or local aligning interactions, there are different phases that have been theoretically analyzed and experimentally observed \cite{vicsek1995novel,marchetti2013hydrodynamics,deblais2018boundaries,giomi2013swarming,dauchot2019dynamics,peruani2018cold}. For example, the full phase diagram of Brownian disks has been discussed \cite{PhysRevLett.121.098003}, the effect of hydrodynamic  interactions in the collective behavior of fish \cite{filella2018model}, the onset of collective and cohesive motion \cite{gregoire2004onset}, the existence of a spontaneously flowing crystal of self-propelled particles \cite{briand2018spontaneously}, to name a few. In these cases, long-range interactions were not explicitly considered \cite{vicsek1995novel,deseigne2010collective,giomi2013swarming,marchetti2013hydrodynamics,peruani2018cold,deblais2018boundaries,PhysRevLett.121.098003,filella2018model,dauchot2019dynamics}. However, recently it has been proposed that dipolar active matter can show a rich phenomenology, where depending on several parameters fission or fusion of particle clusters have been predicted \cite{kaiser2015active,guzman2016fission}, directed self-assembly of quadrupolar particles has been reported \cite{gu2019magnetic}, and in ferromagnetic colloids the collective motion and its different phases experimentally analyzed \cite{kaiser2017flocking}. More broadly, anisotropy alone is already known to reorganize active matter in profound ways. Recent experimental and computational studies of rod-like active particles have shown that shape-induced alignment, steric interactions, and hydrodynamic couplings can generate swarming, flocking, turbulence, clustering, and jamming \cite{Schwarzendahl2022GrowingActiveNematic,CarrilloMora2025Hexbug,Shelke2026ActiveRods,Schueler2026Review}. Our system builds
on that logic by asking how these collective tendencies are modified when the particles
are not only anisotropic and active, but also interact through long-range magnetic forces. The previous results are a clear call to analyze the impact of cluster formation for swimming bacteria that contain a chain of magnetic nano-crystals endowing these organisms with a dipolar moment.  

In what follows, we introduce a general framework to study MAM (Fig.\ref{FIG1}), provide simple methods to measure all parameters of the model (See Supplementary Information S1 and S2), characterize the phases shown by MAM at low densities, and identify the key dimensionless parameters controlling the dynamics. These results provide a calibrated basis for extrapolating the model toward the MTB regime. We observed the evolution of a group of these active particles for several interaction strengths and activities. Using this experimental insight, we implemented a model that accurately reproduces the different dynamical structures observed in experiments (Figs.\ref{FIG1} and \ref{FIG2}): free single particles, quadrupolar dimers, winding of magnetic vortices, unwinding of vortices into ferromagnetic chains, and multipolar clusters (See details of experiments and numerical simulations in Supplementary Information S1-S8).
 We rationalize these findings using dimensionless numbers that depend on the strength of the magnetic interactions between particles, activity, and angular noise (See Supplementary Information and Eqs.\ref{eqMotion}-\ref{eq2}). 
 We explored an enlarged parameter space by using numerical simulations and analytic estimates to understand the main factors that impact in the different phases that MAM display. The phases found in our experiments are relevant for programmed self-assembly, and they should persist across different length scales \cite{kantorovich2013nonmonotonic,whitesides2002self,whitesides2002beyond}. The latter allows us to explore the physical
constraints that may affect key aspects of MTB magnetotaxis using the insight gained from our macroscopic experiments \cite{bellini1963,frankel2009discovery,blakemore1975magnetotactic}.
The system that we study is composed of bacillus-shaped active particles that are built with commercially available  Hexbug Nano robots (HEXBUG Nano, https://www.hexbug.com/nano). We dubbed these macroscopic agents {\bf{MagD-bots}}.Vibration-driven robots of this type have been used previously in active-matter experiments \cite{patterson2017clogging,deblais2018boundaries,dauchot2019dynamics,boudet2021collections,zheng2021self,CarrilloMora2025Hexbug,Baconnier2025SelfAligning}, but their native sharp body geometry, and highly frictional material they are made, makes it unclear that the agreement between experiments and models is real or an artifact. 

To overcome this difficulty, we have dressed the Hexbugs with a 3D printed armor with a bacillus shape. The total body length $L=51.8\times 10^{-3}m$, width $W=20.4\times 10^{-3}m$, mass $M \sim 10 \times 10^{-3}kg$, and simple geometry have allowed us not only to improve the stability of the system by lowering its center of mass, lower the friction between walls and active agents, but also, to simplify numerical simulations. These structured armors allow for the precise positioning of a Neodymium magnet on top of them. Thus, we have built a set of active agents that interact via a magnetic field, and hard-core interactions. We emphasize that in this work the size of magnet is not small compared with the size of the object, making it mandatory to adopt a dumbbell model instead of a point-like dipole \cite{Vokoun201155,Vokoun20093758,ryzhkin2005magnetic,castelnovo2008magnetic,mellado2010dynamics,concha2018PRL,cisternas2021stable}. 
Our main findings are summarized in Fig.\ref{FIG1} where five qualitatively different phases are shown. In the non-magnetic case, phase ${\bf{Free}}$ (Fig.\ref{FIG2}a and Supplementary Movie 1) we observed weak correlations and mainly geometric effects in agreement with previous reports \cite{giomi2013swarming}.  
For large activity and short magnets, we observed particles with weak ferromagnetic correlations that influence the final state, but still without forming any spontaneously self-assembled structure, phase ${\bf{Di}}$ (Fig.\ref{FIG2}b and Supplementary Movie 2). As activity decreases, or as the magnetic interaction strength increases, we observed strong ferromagnetic correlations and the formation of new structures such as dimers, ferromagnetic chains, and vortices not found in the previous cases, phase ${\bf{FM}}$ (Fig.\ref{FIG2} c and Supplementary Movie 3). As magnetic interactions increase, long chains assemble and the system shows clear ferromagnetic order, stable vortices, and the fusion of different objects, phases ${\bf{FM}}$, and ${\bf{V}}$ (Fig.\ref{FIG2} d, Supplementary Movies 4, and 5). 
Finally, for low rotational noise we observe the formation of anti-ferromagnetic ordered clusters that are relevant for the quasi-static self-assembly of magnetic particles, phase ${\bf{C-AF}}$ (Fig. \ref{FIG1})  (See Supplementary Movie 6).

We rationalize our findings considering the Newtonian dynamics of active magnetic matter described by 
\begin{equation}
  \begin{aligned}
M\frac{d \vec{v_{i}}}{dt}  &=F_{0}\hat{n}_{i}-\Gamma \vec{v_{i}} +\vec{f}_{i}^{a} +\vec{f}_{i}^{c}  \\
\label{eqMotion}
I_{z}\frac{d^2\theta_{i}}{dt^2} &= -\Gamma_{r} \frac{d\theta_{i}}{dt}+\left(\vec{\tau}_{i}^{c} +\vec{\tau}_{i}^{a}\right)\cdot \hat{z} +\sqrt{2D_{r}}\eta_{i}(t)
\end{aligned}
\end{equation}

 where $i$ is the particle index, and $\theta_{i}$ is the angle between the horizontal axis and the $\hat{n}_{i}$ director shown in Fig.\ref{FIG1}c. In this model, each rod is represented as a magnetic dumbbell whose two ends carry charges $Q_{\pm}=\pm M_s\pi r^2$. $M$, $F_{0}$, $\Gamma$, $I_{z}$, $\Gamma_{r}$, and $D_{r}$ are the {\bf{MagD-bot}} mass, activity, damping, moment of inertia, rotational damping, and rotational diffusion respectively (See Supplementary Information S1-S3). $\vec{f}^{a}_{i}$, and $\vec{\tau}^{a}_{i}$ represents forces and torques due to contacts. $\vec{f}^{c}_{i}$, $\vec{\tau}^{c}_{i}$ are the force and torque due to Coulomb interactions, and $\eta_{i}(t)$ is a white noise with mean $0$ and variance $1$.

Analyzing the translational equation of motion, the dimensionless numbers $F_{0} \tau (M v_{0})^{-1}$ and $\Gamma\tau M^{-1}$
emerge, where $\tau\sim L/v_{0}$ is a typical time scale for the translational motion of each {\bf{MagD-bot}}. In steady state both terms should be comparable. Therefore, it is enough to consider only one of them to understand the competition between different forcing terms. 
 We called the inverse of the first dimensionless number
\begin{eqnarray}
\alpha =\frac{M v_{0}^{2}}{F_{0} L} .
\label{eq1}
\end{eqnarray}
This dimensionless parameter quantifies the competition between the centripetal force ($M v_{0}/\tau \sim M v_{0}^{2}/L $), and the activity, $F_{0}$, of the {\bf{MagD-bot}}.
 The second dimensionless number relevant in this case appears by considering the rotational dynamics. In the low-mass limit
relevant for MTB, the main forces competing will be the Coulombic torque $\tau^{c} $ and the torsional noise amplitude $\tau^{r}$. We called this parameter:
\begin{eqnarray}
\beta= \frac{\tau^{r}}{\tau^{c}}  
\label{eq2}
\end{eqnarray}
where $\tau^{r} =\left(2 D_{r}\right)^{1/2}$ and $\tau^{c}=\left( \frac{\mu_{o}}{4\pi}\right)\frac{Q^2}{(2R)^2}a $. In this article we focus on two dimensionless parameters ($\alpha$ and $\beta$), but the full dynamics remains richer than this two-parameter projection. Indeed, there are five dimensionless parameters (See Supplementary Information S1, S2).

We proceed to examine the dynamics of the system in detail using molecular dynamics simulations (See Fig.\ref{FIG3} and Supplementary Information). In this description, inertial magnets interact through the full long-range Coulomb potential in the dumbbell approach \cite{Vokoun201155,Vokoun20093758,ryzhkin2005magnetic,castelnovo2008magnetic,mellado2010dynamics,concha2018PRL,cisternas2021stable}. 

 The physical parameters employed in numerical simulations were directly measured (See details in Supplementary Information). The measurement of $F_{0}$, and $\Gamma$ was done by using an incline and analyzing quasi-1D trajectories (See Supplementary Information S4). $I_{z}$ was measured using a torsion pendulum configuration (See Supplementary Information S7), and $\Gamma_{r}$ was measured by analyzing the angular dynamics after a {\bf{MagD-bot}} hits a hard wall (See Supplementary Information S8). Magnetic interactions were measured using special non-magnetic holders installed on an Instron universal testing machine that measured the interaction force between magnets (Supplementary Information S6). From this information we calibrated the magnets used in our experiments obtaining a saturation magnetization as, $M_{s}=(1.1\pm 0.10)\times 10^{6}$ A $m^{-1}$, which is in agreement with the available data for the magnetization of Neodymium rods, validating the dumbbell approximation when the rods are separated by a distance $z>2r$, as previously pointed out \cite{Vokoun201155,Vokoun20093758,mellado2012macroscopic}. That information allowed the measurement of the magnetic charge in the dumbbell approximation.
The ratio between $\Gamma_{r}$ and the rotational noise amplitude, $D_{r}$, was measured using the angular autocorrelation of $\theta(t)$ (See Supplementary Information S8). Therefore, once the independently measured inputs are fixed, the simulations contain no
adjustable fit parameters. Simulation results for {\bf{MagD-bots}} are shown in Fig.\ref{FIG3} in good agreement with experimental results (Fig.\ref{FIG2}). The evolution of three different regimes is shown in Supplementary Movies 7-9. 

We characterized the system by computing the correlations $\langle \hat{n}_{i}\cdot\hat{n}_{j}\rangle$, the magnitude of the average $|\langle \hat{n}_{i}\rangle|$, the magnitude of the average velocity $|\langle \vec{v}_{i}\rangle|$, and average speed $\langle |\vec{v}_{i}|\rangle$ as functions of time \cite{gregoire2004onset,vicsek2012collective,giomi2013swarming}. In Fig.\ref{FIG3} we show representative frames for numerical simulations of MAM where no magnets were present (Fig.\ref{FIG3} a), magnets of length $2a=19.05$ mm (Fig.\ref{FIG3} b), and magnets of length $2a=38.1$ mm (Fig.\ref{FIG3} c). For Fig.\ref{FIG3} a, we found no ferromagnetic order, but from time to time a boundary driven stagnation state \cite{giomi2013swarming}. In Fig.\ref{FIG3} b a signal of ferromagnetic correlations emerges, and finally it is quite clear that ferromagnetic order is present in Fig.\ref{FIG3} c. In agreement with experiments, see Fig.\ref{FIG2} and Supplementary Movies 7-9.

Our macroscopic experiment, and detailed numerical model allow us to draw conclusions about, and make predictions for the phases, and stability of MAM at the microscale. To analyze the impact of increasing magnetic moment in MTB, we have performed detailed numerical simulations in a region of physical parameters relevant for MTB (See Supplementary Tables VIII and IX). Furthermore, the view of the magnetosome chain in MTB as a magnetic dumbbell has been directly measured using extremely sensitive nitrogen-vacancy centers \cite{le2013optical}. For the simulations shown in Fig.\ref{FIG4}, we have used bacterium length $L=5\mu$m, width $W=1.6 \mu$m, magnetic chain length $2a=1.1\mu$m, $Q_{o}\sim 9.4\cdot 10^{-10}$ Am, and average swimming speeds of order $v\sim 30 \mu$m$/$s \cite{khalil2013control,seong2001swimming,schuler1999bacterial,frankel1981magnetotactic,frankel1980navigational,komeili2006magnetosomes}. These values correspond to dimensionless parameters $\beta\sim 10$ and $\alpha \sim 1.5*10^{-7} $, which are far from clustering (see Fig.\ref{FIG1}d). 
To increase the magnetic moment in our in-silico experiments we have increased the charge from $Q_{0}$ to $4Q_{0}$, and then to $10Q_{0}$. We obtained the dynamics of that microscale MAM, Fig. \ref{FIG4} a-c (See Supplementary Movies 10, 11, and 12), and computed the evolution of different order parameters Fig. \ref{FIG4}(d,e). Our data  shows the emergence of a ${\bf{C-AF}}$ phase at large dipolar moment ($\beta<1.0$), together with a sharp decrease in the average speed of the collective motion. This suggests that excessively large magnetic moments would tend to reduce effective motility and compromise magnetotactic performance, which would be detrimental for MTB survival \cite{darwin1964origin}.
For MTB to have an efficient magnetotactic advantage, they should inhabit the {\bf{Free}} phase region of the phase diagram.
That means that in nature the ${\bf{FM}}$,  ${\bf{V}}$, and  ${\bf{C}}$-${\bf{AF}}$ regimes are expected to be disadvantageous because they lower swimming efficiency by interfering with flagellar propulsion \cite{dauparas2016flagellar} and, in some clustered states, by producing compound bodies with reduced net magnetic response and low effective motility \cite{Pan2009HigherFields,Gachon2025Magnetosensing,Schueler2026Review}.

Given that dipolar moment can be increased by increasing $Q$, or by increasing $2a$, we provide a simple bound for $Q_{max}$, the maximum magnetic charge, that a magnetosome chain can have before going into a  ${\bf{C}}$-${\bf{AF}}$ phase. Considering two MTB aligned side by side (Fig.\ref{FIG1}c), they will not form a cluster as long as $\beta>1$. That is, $(\mu_0/2\pi)Q_{max}^{2} a /(2R)^{2}\leq (2D_r)^{1/2} $. This bound gives $Q_{max}\sim 9 \cdot 10^{-10}$Am, for $D_{r}=2.5\cdot 10^{-38} kg^2m^4 s^{-3}$ , $R=0.30\mu$m, and $2a=1.0\mu$m, consistent with observations \cite{dunin1998magnetic,frankel1980navigational,rosenblatt1982birefringence,rosenblatt1985birefringence}. On  the other hand, for fixed magnetic charge, we can find the maximum length for the magnetosome chain to avoid cluster formation. In this case, $a_{max}=(2\pi/\mu_0)(2D_r)^{1/2}(2R)^2/Q^{2}\sim 450$nm in agreement with reported chain lengths \cite{dunin1998magnetic,schuler1999bacterial,frankel1980navigational,komeili2006magnetosomes}. That is, MTB grow magnetic dipoles that are strong enough to align MTB to the geomagnetic field in a noisy environment, but weak enough to avoid the formation of magnetic clusters \cite{Schueler2026Review}. 

Our platform, physical model, and numerical simulations of MAM have allowed us to explore the effects of magnetic interactions in active matter across widely separated length scales, and suggest that increasingly large magnetic moments, while beneficial for alignment in weak fields, can eventually become counterproductive by promoting cluster formation. This identifies a physically grounded mechanism that constrains the useful range of magnetosome-chain magnetic moments in MTB. We anticipate that MAM can be programmed by modifying its magnetic textures, activity, geometry, or the related dimensionless parameters that characterize this system. This provides a step toward functional bioinspired magnetic active matter at all scales.
\begin{acknowledgments}
N.S. and F.G.L. acknowledge ANID (Chile) --Millennium Science Initiative Program-- NCN19 170.
N.S. was partially supported by FONDECYT (Chile) through grant 1180791.
B.G. acknowledges support from ANID PIA/BASAL FB0002.
E.H. was funded by VRIDEI - DICYT project 041931HH.
A.C. acknowledges support from the CODEV Seed Money Program of the \'Ecole Polytechnique F\'ed\'erale de Lausanne (EPFL), the partial support of FONDECYT Regular(Chile) through grants $1210297$ and $1250681$, and the support of the Design Engineering Center at UAI.  
\end{acknowledgments}

\bibstyle{apsrev4-2}
\bibliography{active}

\end{document}